\documentclass[final,5p,times,twocolumn]{elsarticle}

\usepackage{amsmath,amssymb,amsfonts}
\usepackage{graphicx}
\usepackage{booktabs}
\usepackage{multirow}
\usepackage{xcolor}
\usepackage{url}
\usepackage[hidelinks]{hyperref}
\usepackage{tikz}
\usetikzlibrary{arrows.meta, positioning, fit, backgrounds, calc}

\journal{Information systems}   

\begin{document}
\begin{frontmatter}

\title{Controlled Evaluation of Graph and Multimodal Augmentation in RAG for Document Question Answering}

\author[aff1]{Sokipriala Jonah\corref{cor1}}
\ead{jonahs@uni.coventry.ac.uk}
\author[aff1]{Opeoluwa Akinseloyin}
\author[aff1]{Michael Ajao-Olarinoye}
\author[aff1]{Abiola Babatunde}
\cortext[cor1]{Corresponding author}

\affiliation[aff1]{
    organization={Centre for Computational Science and Mathematical Modelling,
    Coventry University},
    city={Coventry},
    country={United Kingdom}
}

\begin{abstract}
Graph and multimodal extensions to retrieval-augmented generation (RAG) are often evaluated end to end, making it difficult to isolate whether gains arise from retrieval, prompt-side context, visual access, generator capability, or benchmark construction. We present a stage- and evidence-controlled evaluation across five RAG configurations, four multimodal generators, and three document corpora. The same LLM-extracted knowledge graph is used either after retrieval as provenance-constrained triple injection (+KG) or during retrieval as entity-bridged passage expansion (+KGret). Prompt-side graph injection yields no consistent accuracy improvement and generally reduces faithfulness. In contrast, +KGret increases gold-evidence completeness from 0.22 to 0.46 on HotpotQA bridge questions and from 0.50 to 0.72 on SPIQA cross-paper questions, improving accuracy for every generator on both evidence-deficient sets while having little effect on retrieval-complete controls. For visual question answering, matched caption-answerable and verified pixel-only protocols show that apparent multimodal gains are sensitive to textual leakage. Programmatic checks reveal answer recoverability from captions, corpus text, and model responses generated without complete gold evidence. Accuracy on incomplete-evidence questions reaches 0.35--0.71 on widely disseminated corpora, compared with 0 on PubLayNet, indicating that raw accuracy can overstate retrieval-attributable performance. These results show that graph augmentation is most effective when it changes retrieval under evidence deficits, while multimodal evaluation requires explicit verification that answers are unavailable through text.
\end{abstract}

\end{frontmatter}



\section{Introduction}

Retrieval-augmented generation (RAG) grounds large language models (LLMs) in
external evidence retrieved at inference time
~\cite{lewis2020rag,gao2023survey}. Most document RAG pipelines nevertheless
reduce documents to extracted text before indexing. This representation is
effective for conventional factual retrieval but can discard information
encoded in figures, tables, spatial structure, and relations distributed across
passages. Optical character recognition can further distort numerical and
layout-sensitive content. Consequently, successful text retrieval does not
necessarily imply that the evidence required for an answer has been recovered
~\cite{yang2024crag,wang2025pixelrag}.

Two extensions address complementary parts of this limitation. Multimodal RAG
retrieves visual content and allows a vision-language model to reason over page
images, figures, and tables~\cite{xia2025mmedrag,faysse2024colpali}. Graph-based
RAG represents entities and relations explicitly so that evidence distributed
across passages or documents can be connected
~\cite{edge2024graphrag,hipporag}. Recent work has expanded both directions to
multi-page, multi-document, and multimodal graph settings
~\cite{cho2024m3docrag,dong2025mmdocrag,wang2026mkgragbench,wang2026multimodalgraphrag}.
However, reported gains remain difficult to compare because the graph may alter
retrieval in one system but only add facts to the generation prompt in another,
visual questions may be answerable from captions or surrounding text, and
widely disseminated benchmark documents may be partially recoverable from model
prior knowledge~\cite{mallen2023popqa,ragvsgraphrag}.

The central methodological problem is therefore one of attribution. An
end-to-end improvement can reflect a better retriever, additional evidence, a
more capable generator, different question construction, or information already
available to the model without the retrieved gold source. This study addresses
that problem by controlling the stage at which graph information is introduced,
verifying the textual recoverability of visual answers, and reporting accuracy
conditioned on whether the complete gold evidence was retrieved. The resulting
design tests not only whether an augmentation helps, but also the mechanism and
evidence conditions under which the gain occurs.

Five configurations are evaluated: a text-only baseline; generation-stage graph
augmentation (+KG), which injects provenance-filtered triples; retrieval-stage
graph augmentation (+KGret), which uses the same graph to expand the candidate
passage set; multimodal augmentation (+multimodal); and a late-fusion graph plus
multimodal configuration (+both). Four vision-capable production generators are
examined across three corpora selected to vary document structure and text
extraction: 1{,}000 PubLayNet-derived scientific pages
~\cite{zhong2019publaynet}, 100 computer-science papers from
SPIQA~\cite{spiqa}, and 2{,}964 Wikipedia paragraphs constructed from
HotpotQA~\cite{yang2018hotpotqa}. Matched controls compare within- and
cross-document reasoning, HotpotQA bridge and comparison questions, and
caption-answerable and pixel-only figure questions.

The study makes three main contributions:
\begin{itemize}
  \item \textbf{Stage-controlled graph evaluation.} The same extracted graph is
  applied at generation time and retrieval time under otherwise fixed
  conditions. This isolates pipeline stage from graph construction and shows
  that graph benefit is concentrated in retrieval, specifically when the base
  retriever leaves an evidence deficit.

  \item \textbf{Verified multimodal benchmark construction.} Figure questions
  are evaluated using matched caption-answerable and pixel-only protocols, with
  programmatic checks for answer recoverability from captions and corpus text.
  The analysis identifies textual leakage and separates it from genuine visual
  interpretation.

  \item \textbf{Evidence-conditioned evaluation across generators and corpora.}
  Accuracy is stratified by gold-evidence completeness, alongside faithfulness,
  relevancy, latency, token use, and cost. This separates improvements
  attributable to retrieved evidence from answers that remain correct when the
  required gold evidence is absent from context.
\end{itemize}

The principal finding is that augmentation value is conditional rather than
universal. Graph structure improves document QA when it is used to recover
missing evidence, but not when it merely reformulates evidence already
retrieved. Visual access is necessary for verified pixel-only questions, yet
its measured benefit depends strongly on how the benchmark excludes textual
answer paths. These results motivate evaluation protocols that report not only
system accuracy, but also where the evidence came from and whether the tested
augmentation changed access to it.

\section{Related Work}

\subsection{RAG Evaluation, Multi-Hop Retrieval, and Model Prior Knowledge}

RAG combines retrieval with generation so that answers can be conditioned on
external evidence rather than model parameters alone
~\cite{lewis2020rag,karpukhin2020dpr}. Retrieval, however, does not guarantee
evidence sufficiency or correct generation. CRAG reports substantial degradation
on difficult and dynamic factual questions~\cite{yang2024crag}, while
HotpotQA, 2WikiMultiHopQA, and MuSiQue expose failures that arise when multiple
passages must be recovered and combined
~\cite{yang2018hotpotqa,ho20202wiki,trivedi2022musique}. Approaches such as
IRCoT and Self-RAG address this problem by coupling retrieval decisions more
closely with reasoning~\cite{trivedi2023ircot,asai2024selfrag}.

Evaluation is further complicated by model prior knowledge. LLMs retain
substantial factual information parametrically~\cite{roberts2020knowledge}, and
recall tends to be higher for popular entities and frequently observed facts
~\cite{mallen2023popqa,kandpal2023longtail}. A correct answer can therefore be
produced even when the complete gold evidence is absent from the retrieved
context. Our evaluation addresses this confound by reporting retrieval metrics
separately from generation metrics and by conditioning accuracy on gold-evidence
completeness. Accuracy, faithfulness, and relevancy are assessed with judge
models outside the generator families, following the RAGAS framework while
acknowledging known limitations of LLM-based evaluation
~\cite{ragas,zheng2023judge}.

\subsection{Multimodal Document RAG and Figure Question Answering}

Multimodal document RAG preserves visual evidence that is lost when documents
are reduced to text. MMed-RAG combines multimodal retrieval with adaptive
context selection~\cite{xia2025mmedrag}; PixelRAG retrieves and reads document
screenshots directly~\cite{wang2025pixelrag}; and ColPali represents pages in a
vision-language embedding space~\cite{faysse2024colpali}. M3DocRAG and
MMDocRAG extend evaluation to multi-page and multi-document settings and to
broader model comparisons~\cite{cho2024m3docrag,dong2025mmdocrag}. Related
benchmarks such as DocVQA, ChartQA, and SPIQA evaluate question answering over
document images, charts, figures, and tables
~\cite{mathew2021docvqa,masry2022chartqa,spiqa}.

These benchmarks establish the importance of visual evidence but do not remove
a key construction issue: a question intended to require an image may still be
answerable from a caption, surrounding prose, or other retrievable text. In this
study, caption-answerable and pixel-only questions are treated as distinct
evaluation regimes, and pixel-only status is verified against the actual text
available to the retriever. CLIP~\cite{radford2021clip} is used as a controlled
visual retriever so that benchmark construction, retrieval quality, and
figure-reading capability can be analysed separately.

\subsection{Graph-Augmented RAG and Pipeline Stage}

GraphRAG uses LLM-extracted entity structure to support corpus-level reasoning
and query-focused summarisation~\cite{edge2024graphrag}. Subsequent systems
explore lightweight graph indexing, hierarchical summarisation, graph traversal,
and graph-conditioned prompting
~\cite{guo2024lightrag,sarthi2024raptor,sun2024thinkongraph,jiang2023structgpt,baek2023kaping,peng2024graphragsurvey}.
Multimodal variants increasingly place visual and textual information in a
shared graph representation~\cite{wang2026mkgragbench,wang2026multimodalgraphrag}.

The stage at which graph structure is applied is especially important.
HippoRAG uses entity associations as part of retrieval, allowing graph structure
to recover passages that embedding similarity misses~\cite{hipporag}. In
contrast, prompt-side graph augmentation can only help if reformulating
already-retrieved evidence makes it easier for the generator to use. Systematic
comparisons report no universal advantage for graph-based retrieval and show
that effects depend on question type and retrieval demand
~\cite{ragvsgraphrag}. The present study isolates this distinction directly by
applying the same extracted graph at both stages with fixed questions, indices,
and generators. The comparison therefore tests whether graph value derives from
structured evidence presentation or from access to previously unretrieved
evidence.

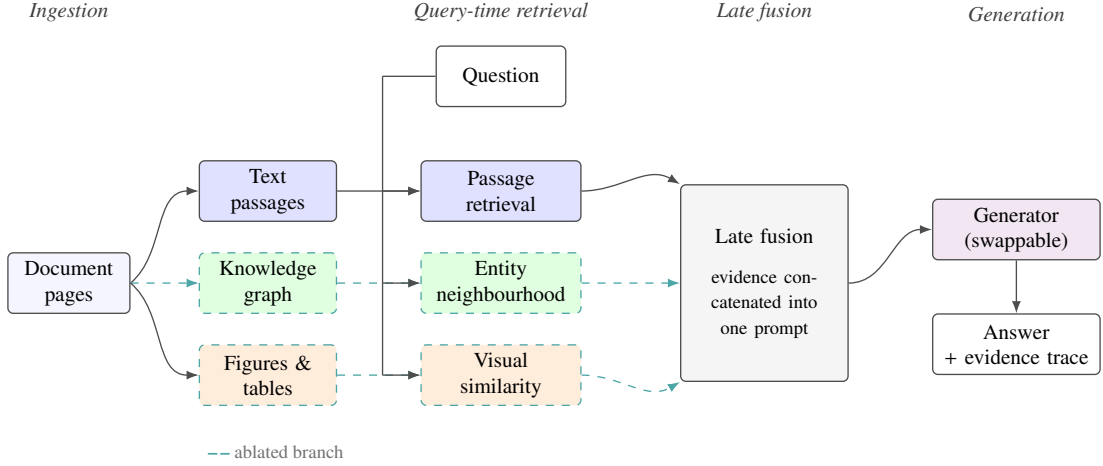
\begin{figure*}[h]
\centering
\begin{tikzpicture}[
  font=\footnotesize,
  node distance=6mm and 9mm,
  base/.style   = {draw=black!70, rounded corners=2pt, align=center,
                   minimum height=8mm, inner sep=3pt, line width=0.5pt},
  ing/.style    = {base, fill=blue!4},
  txt/.style    = {base, fill=blue!12},
  kg/.style     = {base, fill=green!12, draw=teal!70, dashed},
  vis/.style    = {base, fill=orange!14, draw=teal!70, dashed},
  fuse/.style   = {base, fill=black!4, minimum height=26mm, text width=20mm},
  gen/.style    = {base, fill=violet!10},
  term/.style   = {base, fill=white},
  ar/.style     = {-{Latex[length=1.6mm]}, draw=black!70, line width=0.5pt},
  ard/.style    = {-{Latex[length=1.6mm]}, draw=teal!70, line width=0.5pt, dashed},
  lane/.style   = {font=\footnotesize\itshape, text=black!75},
]

\node[ing, text width=14mm] (doc) {Document\\pages};

\node[txt, text width=16mm, above right=4mm and 9mm of doc] (pass) {Text\\passages};
\node[kg,  text width=16mm, right=9mm of doc]               (graph) {Knowledge\\graph};
\node[vis, text width=16mm, below right=4mm and 9mm of doc] (figs)  {Figures \&\\tables};

\draw[ar]  (doc.east) to[out=40,  in=180] (pass.west);
\draw[ard] (doc.east) -- (graph.west);
\draw[ar]  (doc.east) to[out=-40, in=180] (figs.west);

\node[txt, text width=19mm, right=11mm of pass]  (rpass)  {Passage\\retrieval};
\node[kg,  text width=19mm, right=11mm of graph] (rgraph) {Entity\\neighbourhood};
\node[vis, text width=19mm, right=11mm of figs]  (rvis)   {Visual\\similarity};

\draw[ar]  (pass)  -- (rpass);
\draw[ard] (graph) -- (rgraph);
\draw[ard] (figs)  -- (rvis);

\node[term, text width=15mm, above=7mm of rpass] (q) {Question};
\coordinate (qdrop) at ($(rpass.west) + (-5mm, 0)$);
\coordinate (qtop)  at (qdrop |- q);
\draw[draw=black!70, line width=0.5pt] (q.west) -- (qtop);
\draw[draw=black!70, line width=0.5pt] (qtop) -- ($(rvis.west) + (-5mm, 0)$);
\draw[ar] ($(rpass.west)  + (-5mm, 0)$) -- (rpass.west);
\draw[ar] ($(rgraph.west) + (-5mm, 0)$) -- (rgraph.west);
\draw[ar] ($(rvis.west)   + (-5mm, 0)$) -- (rvis.west);

\node[fuse, right=13mm of rgraph] (fusion)
  {Late fusion\\[2mm]\scriptsize evidence concatenated into one prompt};

\draw[ar]  (rpass.east)  to[out=0, in=150] (fusion.north west);
\draw[ard] (rgraph.east) -- (fusion.west);
\draw[ard] (rvis.east)   to[out=0, in=210] (fusion.south west);

\node[gen, text width=20mm, above right=3mm and 11mm of fusion.east] (gen)
  {Generator\\(swappable)};
\node[term, text width=20mm, below=7mm of gen] (ans) {Answer\\+ evidence trace};

\draw[ar] (fusion.east) to[out=0, in=180] (gen.west);
\draw[ar] (gen) -- (ans);

\coordinate (lanerow) at ($(q.north) + (0, 2mm)$);
\node[lane, anchor=south] at (doc      |- lanerow) {Ingestion};
\node[lane, anchor=south] at (q        |- lanerow) {Query-time retrieval};
\node[lane, anchor=south] at (fusion   |- lanerow) {Late fusion};
\node[lane, anchor=south] at (gen      |- lanerow) {Generation};

\node[font=\scriptsize, text=black!55, align=left, below=4mm of figs.south west,
      anchor=north west] (key)
  {\textcolor{teal!70}{\textbf{--\,--}}~ablated branch};

\end{tikzpicture}
\caption{Stage-controlled multimodal graph-RAG architecture. Text, graph, and visual evidence are indexed separately. The four late-fusion conditions ablate prompt-side graph and visual evidence while preserving text retrieval. The +KGret condition reuses the same graph to expand the text candidate set before generation, enabling direct comparison between generation-stage and retrieval-stage graph augmentation.}
\label{fig:architecture}
\end{figure*}

\section{Method}

The framework represents each corpus through text passages, visual regions
where available, and an LLM-extracted knowledge graph. The core experimental
principle is to keep these representations fixed while changing only how they
are exposed to the generator. Figure~\ref{fig:architecture} summarises the
architecture.

\subsection{Problem Formulation}
\label{sec:formulation}

Let $\mathcal{C}$ be the text passages, $\mathcal{V}$ the figures and tables,
and $\mathcal{G}=(\mathcal{N},\mathcal{E})$ a directed graph whose edges are
subject--relation--object triples extracted from $\mathcal{C}$. Each item
retains a provenance identifier $\pi(\cdot)$.

For query $q$, text, visual, and generation-stage graph evidence are defined as
\begin{align}
R_\text{text}(q) &= \operatorname{TopK}_{c \in \mathcal{C}}
\cos\!\left(\phi_t(q),\phi_t(c)\right), \\
R_\text{vis}(q) &= \operatorname{TopM}_{v \in \mathcal{V}}
\cos\!\left(\phi_v(q),\phi_v(v)\right), \\
R_\text{kg}(q) &= \left\{e \in \mathcal{E}: \eta(e) \sqsubseteq q,
\ \pi(e) \in \pi\!\left(R_\text{text}(q)\right)\right\},
\end{align}
where $\phi_t$ is the text encoder, $\phi_v$ maps text and images into a shared
space, $\eta(e)$ denotes a matched entity, and $\sqsubseteq$ denotes whole-word
matching. The provenance constraint prevents prompt-side graph facts from
unrelated retrieved units.

With $\oplus$ denoting prompt concatenation and $g_\theta$ the generator, the
four late-fusion configurations are
\begin{equation}
A(q;\alpha,\beta)=g_\theta\!\left(
q \oplus R_\text{text}(q)
\oplus \alpha R_\text{kg}(q)
\oplus \beta\sigma\!\left(R_\text{vis}(q)\right)
\right),
\end{equation}
where $\alpha,\beta\in\{0,1\}$ and $\sigma$ renders visual evidence as pixels
or captions. These settings produce the baseline, +KG, +multimodal, and +both
conditions.

The fifth condition, \textbf{+KGret}, applies the same graph during retrieval.
Let $\mathcal{E}(q)$ be the matched graph nodes and let
$\mathrm{pages}(\mathcal{E}(q))$ be the provenance units connected to those
nodes. Candidate passages are expanded as
\begin{equation}
\label{eq:kgret}
R_\text{kgret}(q)=R_\text{text}(q)\cup
\operatorname{Top}_{B}\!\left\{c\in\mathcal{C}:
\pi(c)\in\mathrm{pages}(\mathcal{E}(q))\setminus
\pi\!\left(R_\text{text}(q)\right)\right\},
\end{equation}
with $B=2$ bridged chunks ranked by query similarity. No triples are injected
in +KGret. Thus +KG can only restructure evidence already retrieved, whereas
+KGret can only add previously unretrieved passages. This separation is the key
stage control used in the experiments.
\label{sec:kgret}

\subsection{Corpora and Preprocessing}
\label{sec:corpora}

Three corpora were selected to provide different evidence structures and text
extraction conditions.

\subsubsection{PubLayNet}
The visually rich corpus contains 1{,}000 pages from the
\texttt{small-publaynet-wds} distribution derived from
PubLayNet~\cite{lhoestq_small_publaynet_wds,zhong2019publaynet}. Regions
labelled text, title, list, or table are processed with Tesseract OCR and merged
into page-level text. Figure and table regions are retained as image crops, so
tables are available through both OCR text and pixels. Each crop is captioned
once with GPT-4o using a fixed concise-description prompt. OCR text is segmented
into 500-character windows with 50-character overlap. Text chunks use
\texttt{text-embedding-3-small}~\cite{openai2024embeddings}; visual crops use
CLIP ViT-B/32~\cite{radford2021clip}; both are indexed with FAISS under cosine
similarity~\cite{johnson2019faiss}.

\subsubsection{SPIQA}
The second corpus contains 100 papers from the manually curated test-A split of
SPIQA~\cite{spiqa}, comprising 8{,}474 text chunks and 1{,}126 figure/table
images. Text is derived from the distributed TeX paragraphs rather than OCR,
and the extracted graph contains 50{,}123 triples. Recurring terminology across
papers provides the cross-document entity structure used in the cross-paper
question set. Images are indexed with the same CLIP model and captioned with the
same ingestion prompt used for PubLayNet.

\subsubsection{HotpotQA}
The text-only corpus pools the context paragraphs of 300 HotpotQA distractor
validation questions~\cite{yang2018hotpotqa}, yielding 2{,}964 unique Wikipedia
paragraphs, 5{,}128 chunks, and 36{,}447 extracted triples. One provenance unit
corresponds to one Wikipedia article. Only the baseline and graph conditions
are evaluated because the corpus has no visual channel.

Across corpora, text chunking, embedding, indexing, and triple extraction follow
the same processing contract. The provenance unit is a page for PubLayNet, a
paper for SPIQA, and an article for HotpotQA.

\subsection{Retrieval and Evidence Fusion}

The baseline retrieves the top $k=3$ text chunks and instructs the generator to
answer only from the supplied evidence or abstain. GPT-4o-mini extracts graph triples once per text chunk, which
are cached in a directed NetworkX graph~\cite{hagberg2008networkx}. Generation-
stage graph matching excludes entities shorter than five characters and admits
at most eight neighbouring triples whose provenance is already represented in
text retrieval. Retrieval-stage graph augmentation uses the same entity and
length rules but appends at most two previously unretrieved chunks from
entity-linked provenance units.

For multimodal retrieval, CLIP ranks figures and tables by text-image similarity.
Retrieval is evaluated to depth $m=5$, while the generator receives either the
top-ranked crop as pixels or the captions of the top three crops. Pixel input is
used only for figure questions; text and multi-hop questions receive captions in
multimodal conditions. This gating prevents irrelevant images from being added
to queries with no visual evidence requirement.
\label{sec:kg}
\label{sec:multimodal}

\subsection{Explainability}

Each response log stores the retrieved text and provenance, graph facts or
bridged chunks, visual retrieval scores, and the evidence supplied to the
generator. These traces support per-question error analysis and allow every
aggregate result to be linked to its underlying evidence. Explainability is
therefore used as an audit mechanism rather than as a separately scored outcome.

\section{Experiments and Results}
\label{sec:experiments}

\subsection{Experimental Design}
\label{sec:design}

The evaluation controls three factors: evidence configuration, generator, and
corpus. Five configurations are considered where applicable: baseline, +KG,
+KGret, +multimodal, and +both. Four production generators are tested, with
retrieval indices, graph caches, and retrieval parameters fixed within each
corpus. Matched controls are used to distinguish architectural effects from
question difficulty: SPIQA within-paper versus cross-paper multi-hop questions,
HotpotQA bridge versus comparison questions, and PubLayNet caption-answerable
versus pixel-only figure questions. +KGret is emphasised on question sets with
an observable evidence deficit because candidate expansion has no expected
benefit when the base retriever is already complete.

\subsection{Question Sets}
\label{sec:qsets}

PubLayNet contributes 35 single-passage, 30 within-page multi-hop, 35 pixel-only
figure, and 35 caption-answerable figure questions. SPIQA contributes 35
single-passage, 30 within-paper multi-hop, 50 cross-paper multi-hop, and 35
figure questions. HotpotQA contributes 50 bridge and 50 comparison questions
with two gold source paragraphs each.

Question authorship is separated from the evaluated generators. DeepSeek-Chat
authors text and multi-hop questions, and Claude Haiku~4.5 authors figure
questions, both with fixed seeds and temperature 0. Programmatic constraints
remove modality cues, source-document references, archive identifiers, and
underspecified wording; all sets are subsequently reviewed manually. SPIQA
cross-paper questions are seeded from graph entities that occur in at least two
papers and contain the linking entity explicitly, ensuring that the graph covers
the intended relation. HotpotQA bridge and comparison labels are retained from
the benchmark.

\subsubsection{Figure question integrity verification}
\label{sec:integrity}

Caption-answerable questions are authored from the pipeline caption, whereas
pixel-only questions are authored from the crop itself. Because construction
intent does not guarantee that the answer is unavailable through text, each
pixel-only item is checked programmatically against its crop caption and the
retrievable text of its source document. PubLayNet contains no caption matches
in the pixel-only set (0/35). On SPIQA, the same checks yield 5
caption-contaminated items, 4 text-recoverable items, and 26 strict-pixel items.
Headline visual claims use the verified tiers rather than the authoring label
alone.

\subsection{Generators and Implementation}
\label{sec:generators}

\begin{table}[t]
\caption{Generators used in the evaluation. Prices are in USD per million
input and output tokens at the time of the experiments.}
\label{tab:models}
\centering
\footnotesize
\begin{tabular}{@{}llcr@{\,/\,}l@{}}
\hline
Model & Host & Access & \multicolumn{2}{c}{\$/1M in/out} \\
\hline
GPT-4o-mini             & OpenAI    & Closed & 0.15 & 0.60 \\
Gemini 3.1 Flash-Lite   & Google    & Closed & 0.10 & 0.40 \\
Llama 4 Scout           & DeepInfra & Open   & 0.08 & 0.30 \\
Llama 4 Maverick        & DeepInfra & Open   & 0.15 & 0.60 \\
\hline
GPT-4o (diagnostic)     & OpenAI    & Closed & 2.50 & 10.00 \\
\hline
\end{tabular}
\end{table}

The production comparison includes GPT-4o-mini, Gemini~3.1~Flash-Lite,
Llama~4~Scout, and Llama~4~Maverick; GPT-4o is used only as a higher-cost
diagnostic. Models are evaluated in their served configurations, including FP8
quantisation for Maverick and billed reasoning tokens for Gemini. The open-
weight models are accessed through hosted DeepInfra endpoints, so the access
comparison concerns weights and serving behaviour rather than local deployment.
A common client records input/output tokens, latency, failures, and cost for
every generation and judging call. Aggregate tables and figures are regenerated
from the released per-question logs.

\subsection{Metrics}
\label{sec:metrics}

Let $S_q$ be the set of gold provenance units for question $q$ and
$R_q=(r_1,\ldots,r_k)$ the ranked retrieval results. Retrieval is measured by
\begin{align}
\mathrm{Recall@}k &= \frac{1}{|Q|}\sum_{q\in Q}
\mathbb{1}\!\left[S_q\cap R_q\neq\varnothing\right],\\
\mathrm{MRR} &= \frac{1}{|Q|}\sum_{q\in Q}
\frac{1}{\min\{i:r_i\in S_q\}},\\
\mathrm{Completeness} &= \frac{1}{|Q|}\sum_{q\in Q}
\mathbb{1}\!\left[S_q\subseteq R_q\right],
\end{align}
where the MRR term is zero if no gold source is retrieved. Completeness is
reported for multi-source questions because answering them requires recovery of
the full gold set. Figure questions are scored against the visual ranking and
text questions against the text ranking.

Generation quality is measured by binary accuracy, faithfulness, and relevancy
judgements following RAGAS~\cite{ragas}. DeepSeek-Chat judges text outputs and
Claude Haiku~4.5 judges figure faithfulness; neither belongs to an evaluated
generator family. Accuracy is also conditioned on evidence completeness. The
incomplete-evidence stratum is reported as non-retrieval-attributable accuracy,
because a correct answer in this stratum cannot be credited to complete gold
evidence in the prompt.

\subsection{Four-Way Late-Fusion Ablation}
\label{sec:ablation}

\begin{table*}[t]
\caption{Four-way ablation across four generators and three question types on
the 1{,}000-page PubLayNet corpus. Figure results use the pixel-only set. Best
accuracy for each generator and question type is shown in bold.}
\label{tab:ablation}
\centering
\footnotesize
\setlength{\tabcolsep}{4pt}
\resizebox{\textwidth}{!}{%
\begin{tabular}{ll|ccc|ccc|ccc}
\hline
 & & \multicolumn{3}{c|}{\textbf{Text} ($n=35$)} & \multicolumn{3}{c|}{\textbf{Multi-hop} ($n=30$)} & \multicolumn{3}{c}{\textbf{Figure, pixel-only} ($n=35$)} \\
Generator & System & Acc. & Faith. & Rel. & Acc. & Faith. & Rel. & Acc. & Faith. & Rel. \\
\hline
\multirow{4}{*}{GPT-4o-mini}
 & baseline     & \textbf{0.657} & 0.714 & 0.800 & 0.433 & 0.500 & 0.633 & 0.000 & 0.000 & 0.000 \\
 & +KG          & 0.571 & 0.571 & 0.743 & 0.367 & 0.467 & 0.600 & 0.000 & 0.000 & 0.000 \\
 & +multimodal  & \textbf{0.657} & 0.714 & 0.829 & 0.433 & 0.500 & 0.633 & 0.086 & 0.057 & 0.200 \\
 & +both        & 0.600 & 0.629 & 0.800 & \textbf{0.467} & 0.467 & 0.667 & \textbf{0.114} & 0.114 & 0.257 \\
\hline
\multirow{4}{*}{Gemini 3.1 Flash-Lite}
 & baseline     & 0.514 & 0.600 & 0.771 & 0.600 & 0.467 & 0.633 & 0.000 & 0.000 & 0.029 \\
 & +KG          & \textbf{0.600} & 0.543 & 0.771 & 0.533 & 0.433 & 0.733 & 0.000 & 0.000 & 0.086 \\
 & +multimodal  & 0.571 & 0.600 & 0.800 & \textbf{0.633} & 0.500 & 0.867 & 0.057 & 0.171 & 0.600 \\
 & +both        & \textbf{0.600} & 0.571 & 0.800 & 0.567 & 0.467 & 0.767 & \textbf{0.086} & 0.086 & 0.286 \\
\hline
\multirow{4}{*}{Llama 4 Scout}
 & baseline     & 0.543 & 0.686 & 0.800 & 0.500 & 0.467 & 0.700 & 0.000 & 0.000 & 0.114 \\
 & +KG          & \textbf{0.600} & 0.571 & 0.829 & \textbf{0.533} & 0.467 & 0.667 & 0.000 & 0.000 & 0.086 \\
 & +multimodal  & \textbf{0.600} & 0.571 & 0.771 & 0.500 & 0.533 & 0.733 & \textbf{0.114} & 0.114 & 0.429 \\
 & +both        & 0.571 & 0.543 & 0.800 & \textbf{0.533} & 0.500 & 0.600 & 0.057 & 0.086 & 0.457 \\
\hline
\multirow{4}{*}{Llama 4 Maverick}
 & baseline     & 0.657 & 0.571 & 0.857 & 0.500 & 0.467 & 0.733 & 0.000 & 0.000 & 0.371 \\
 & +KG          & \textbf{0.714} & 0.543 & 0.857 & 0.533 & 0.400 & 0.667 & 0.000 & 0.000 & 0.200 \\
 & +multimodal  & 0.600 & 0.629 & 0.857 & \textbf{0.567} & 0.433 & 0.867 & \textbf{0.086} & 0.114 & 0.600 \\
 & +both        & 0.657 & 0.486 & 0.857 & 0.533 & 0.367 & 0.733 & \textbf{0.086} & 0.171 & 0.514 \\
\hline
\end{tabular}}
\end{table*}

On PubLayNet, text retrieval is identical across the four late-fusion systems:
Recall@3 is 0.914 (MRR 0.886) for text questions and 0.900 (MRR 0.883) for
multi-hop questions. No late-fusion configuration consistently improves text or
multi-hop accuracy across generators. Generation-stage graph augmentation
changes mean accuracy by only $+0.028$ on text, $-0.017$ on multi-hop, and
$0.000$ on figure questions, with mixed signs across models. Because +KG is
provenance constrained, it mainly reformulates evidence already present in the
retrieved passages and adds little new information.

Verified pixel-only questions show a different pattern. Both text-only
conditions score 0 for every generator, while multimodal access raises accuracy
to 0.057--0.114. Visual evidence is therefore necessary on this set, but the
small absolute gain indicates that visual retrieval and figure reading remain
substantial bottlenecks.

\subsection{Pipeline Stage Determines the Value of Graph Augmentation}
\label{sec:stage}

\begin{table*}[t]
\centering
\caption{Baseline, generation-stage (+KG), and retrieval-stage (+KGret) graph
augmentation across four generators. Cells show accuracy (faithfulness).
Completeness is the fraction of questions whose full gold evidence appears in
the candidate set; it is identical across generators within a system because
the retrieval stack is fixed, and differs for +KGret by design.}
\label{tab:threeway}
\footnotesize
\setlength{\tabcolsep}{4pt}
\resizebox{\textwidth}{!}{%
\begin{tabular}{@{}llccccc@{}}
\toprule
Question set & System & Complete. & GPT-4o-mini & Gemini 3.1 F-L & Llama 4 Scout & Llama 4 Maverick \\
\midrule
SPIQA cross-paper   & baseline & 0.50 & 0.620 (0.660) & 0.820 (0.800) & 0.780 (0.720) & 0.700 (0.580) \\
($n{=}50$)          & +KG      & 0.50 & 0.700 (0.640) & 0.780 (0.780) & 0.820 (0.800) & 0.760 (0.700) \\
                    & +KGret   & 0.72 & \textbf{0.900} (0.860) & \textbf{0.880} (0.900) & \textbf{0.820} (0.760) & \textbf{0.840} (0.720) \\
\midrule
HotpotQA bridge     & baseline & 0.22 & 0.520 (0.480) & 0.340 (0.340) & 0.520 (0.380) & 0.560 (0.540) \\
($n{=}50$)          & +KG      & 0.22 & 0.480 (0.400) & 0.400 (0.360) & 0.540 (0.340) & 0.600 (0.400) \\
                    & +KGret   & 0.46 & \textbf{0.720} (0.640) & \textbf{0.560} (0.580) & \textbf{0.660} (0.400) & \textbf{0.680} (0.680) \\
\midrule
HotpotQA comparison & baseline & 0.86 & 0.880 (0.760) & 0.920 (0.840) & 0.960 (0.900) & 0.920 (0.840) \\
(control, $n{=}50$) & +KG      & 0.86 & 0.900 (0.700) & 0.900 (0.760) & 0.920 (0.720) & 0.920 (0.700) \\
                    & +KGret   & 0.98 & 0.880 (0.840) & 0.940 (0.820) & \textbf{0.980} (0.880) & 0.920 (0.900) \\
\midrule
SPIQA within-paper  & baseline & 1.00 & 0.767 (0.833) & 0.767 (0.833) & 0.667 (0.767) & 0.767 (0.767) \\
($n{=}30$)          & +KG      & 1.00 & 0.700 (0.767) & 0.767 (0.833) & 0.800 (0.767) & 0.733 (0.800) \\
\bottomrule
\end{tabular}}
\end{table*}

Generation-stage +KG remains inconsistent on the evidence-deficient sets. On
SPIQA cross-paper questions its accuracy change ranges from $-0.040$ to
$+0.080$ across generators; on HotpotQA bridge questions the range is
$-0.040$ to $+0.060$. Faithfulness also declines in most +KG comparisons.

Retrieval-stage +KGret produces a qualitatively different result. SPIQA
cross-paper completeness increases from 0.50 to 0.72, with accuracy gains of
$+0.280$, $+0.060$, $+0.040$, and $+0.140$ across the four generators.
HotpotQA bridge completeness increases from 0.22 to 0.46, with gains of
$+0.200$, $+0.220$, $+0.140$, and $+0.120$ (mean $+0.170$). Thus the same
graph improves every generator when it expands retrieval, whereas prompt-side
triple injection does not yield a generator-independent benefit.

\begin{figure*}[t]
\centering
\includegraphics[width=0.9\textwidth]{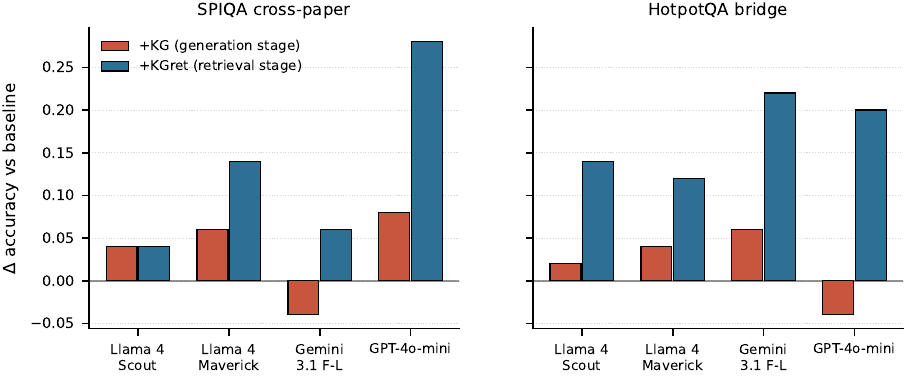}
\caption{Per-generator change in accuracy over the baseline for
generation-stage (+KG) and retrieval-stage (+KGret) graph augmentation on the
two evidence-deficient question sets. Generation-stage effects vary in sign
across generators; retrieval-stage effects are positive for every generator.}
\label{fig:stage}
\end{figure*}

\subsection{Evidence Deficit Explains the Retrieval-Stage Gain}
\label{sec:mechanism}

HotpotQA provides a mechanism control. Bridge questions require a second
paragraph reachable through an entity association, whereas comparison questions
name both target entities and are usually complete under text retrieval alone.
If +KGret benefited merely from longer context, both question types should
improve similarly.

\begin{table}[t]
\centering
\caption{Mechanism control on HotpotQA. Means over the four production
generators. Bridge questions exhibit the evidence deficit (completeness 0.22)
that retrieval-stage augmentation exists to repair; comparison questions do
not.}
\label{tab:mechanism}
\begin{tabular}{@{}llcccc@{}}
\toprule
Set & System & Complete. & Acc. & Faith. & $\Delta$Acc. \\
\midrule
Bridge     & baseline & 0.22 & 0.485 & 0.435 & \textit{n/a} \\
           & +KG      & 0.22 & 0.505 & 0.375 & $+0.020$ \\
           & +KGret   & 0.46 & 0.655 & 0.575 & $\mathbf{+0.170}$ \\
\midrule
Comparison & baseline & 0.86 & 0.920 & 0.835 & \textit{n/a} \\
           & +KG      & 0.86 & 0.910 & 0.720 & $-0.010$ \\
           & +KGret   & 0.98 & 0.930 & 0.860 & $+0.010$ \\
\bottomrule
\end{tabular}
\end{table}

The observed pattern supports evidence recovery rather than generic context
expansion. Mean accuracy increases by $0.170$ on bridge questions but only
$0.010$ on comparison questions, although expansion fires on almost every item
in both sets. One-hop entity-to-provenance association accounts for nearly all
of the completeness gain; two-hop expansion recovers only one additional bridge
question. Residual failures are attributable to the two-chunk budget and to
entity surface-form mismatches, indicating that candidate allocation and entity
canonicalisation are the main limits of the current mechanism.

\subsection{Figure Question Construction Changes the Apparent Multimodal Gain}
\label{sec:construction}

\begin{table}
\caption{Comparison of caption-answerable and pixel-only figure questions. The
two sets use the same corpus, retriever, and production generators.}
\label{tab:ceiling}
\centering
\footnotesize
\begin{tabular}{lcc}
\hline
 & Caption-answerable & Pixel-only \\
 & ($n=35$) & ($n=35$) \\
\hline
Answer in caption            & 25/34 (74\%) & 0/35 (0\%) \\
Cell-value answers           & 1/35 (3\%)   & 23/35 (66\%) \\
Mean answer length (chars)   & 60           & 28 \\
\hline
\multicolumn{3}{l}{\textbf{Text-only baseline accuracy}} \\
\hline
GPT-4o-mini                  & 0.200 & 0.000 \\
Gemini 3.1 Flash-Lite        & 0.200 & 0.000 \\
Llama 4 Scout                & 0.143 & 0.000 \\
Llama 4 Maverick             & 0.171 & 0.000 \\
GPT-4o (diagnostic)          & --    & 0.000 \\
\hline
\multicolumn{3}{l}{\textbf{Multimodal accuracy}} \\
\hline
GPT-4o-mini                  & 0.371 & 0.086 \\
Gemini 3.1 Flash-Lite        & 0.314 & 0.057 \\
Llama 4 Scout                & 0.257 & 0.114 \\
Llama 4 Maverick             & 0.257 & 0.086 \\
\hline
\end{tabular}
\end{table}

On PubLayNet, 25 of 34 available caption-answerable references occur directly
in the corresponding pipeline caption, whereas none of the 35 pixel-only
answers do. Text-only accuracy is consequently 0.143--0.200 on
caption-answerable questions but 0 on the pixel-only set for every production
generator. The multimodal systems also score higher on caption-answerable items
(0.257--0.371) than on verified pixel-only items (0.057--0.114). A benchmark
constructed from captions therefore mixes caption recovery with visual
interpretation and can overstate the apparent value of multimodal retrieval.

\begin{table}[t]
\centering
\caption{Figure-question integrity tiers on SPIQA ($n{=}35$; correct counts
per tier, all four generators). The tiers are determined by programmatic
verification of answer absence from the caption and from the retrievable
corpus text.}
\label{tab:integrity}
\begin{tabular}{@{}lcccc@{}}
\toprule
 & \multicolumn{2}{c}{Caption-cont.\ ($n{=}5$)} & \multicolumn{2}{c}{Strict-pixel ($n{=}26$)} \\
Generator & text-only & +mm & text-only & +mm \\
\midrule
GPT-4o-mini          & 2/5 & 2/5 & 2/26 & 2/26 \\
Gemini 3.1 F-L       & 2/5 & 2/5 & 3/26 & 0/26 \\
Llama 4 Scout        & 2/5 & 2/5 & 3/26 & 1/26 \\
Llama 4 Maverick     & 2/5 & 1/5 & 2/26 & 2/26 \\
\bottomrule
\end{tabular}
\end{table}

SPIQA demonstrates that caption leakage is not the only textual path. The 35
figure questions include 5 caption-contaminated items, 4 additional items whose
answers occur in retrievable paper text, and 26 strict-pixel items. Even within
the strict-pixel tier, text-only systems answer 2--3 of 26 questions correctly,
which is consistent with model prior knowledge rather than retrieved gold
evidence. Multimodal augmentation does not improve the strict-pixel tier. The
integrity of a visual QA benchmark therefore depends on the authoring protocol,
the corpus text-extraction route, and the extent to which the answer can be
produced without complete retrieved evidence.

\subsection{Visual Retrieval and Figure Reading Remain Bottlenecks}
\label{sec:visual}

\begin{table}
\caption{Higher-cost diagnostic on pixel-only figure questions.}
\label{tab:diagnostic}
\centering
\footnotesize
\begin{tabular}{llccc}
\hline
Generator & System & Acc. & Faith. & Rel. \\
\hline
\multirow{2}{*}{GPT-4o-mini}
 & baseline    & 0.000 & 0.000 & 0.000 \\
 & +multimodal & 0.086 & 0.057 & 0.200 \\
\hline
\multirow{2}{*}{GPT-4o}
 & baseline    & 0.000 & 0.000 & 0.000 \\
 & +multimodal & 0.143 & 0.171 & 0.886 \\
\hline
\end{tabular}
\end{table}

\begin{figure}
\centering
\includegraphics[width=\columnwidth]{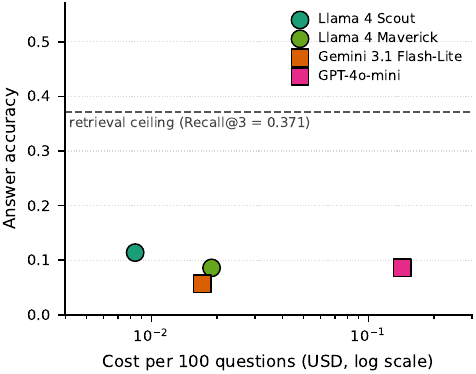}
\caption{Cost and accuracy on pixel-only figure questions for the multimodal
system. The horizontal line shows Recall@3 of 0.371. Because only the top-ranked
crop is passed to the generator, Recall@1 of 0.229 is the operational retrieval
ceiling.}
\label{fig:cost_accuracy}
\end{figure}

For PubLayNet pixel-only questions, image Recall@3 is 0.371 and MRR is 0.286,
but only the top-ranked crop is passed to the generator. Recall@1 is therefore
the operational retrieval ceiling at 0.229, compared with production accuracy
of 0.057--0.114. The gap identifies two separate errors: the gold crop is often
not ranked first, and the generator may still fail to extract the required value
when the correct crop is supplied. This is consistent with the use of a global
CLIP embedding for questions that often target fine-grained numerical cells,
labels, or axes.

The higher-cost GPT-4o diagnostic raises multimodal accuracy from 0.086 to
0.143 relative to GPT-4o-mini, equivalent to two additional correct answers.
The limited gain indicates that generator scale alone does not remove the
retrieval and visual-reading constraints.

\subsection{Evidence-Conditioned Accuracy Reveals a Non-Retrieval Floor}
\label{sec:leakage}

\begin{table}[t]
\centering
\caption{Accuracy conditioned on gold-evidence completeness, pooled over the
four generators. Baseline accuracy on incomplete-evidence questions is a
parametric floor: correct answers produced without the required evidence in
context.}
\label{tab:conditioned}
\begin{tabular}{@{}llcc@{}}
\toprule
Question set & System & Acc.$\mid$compl. & Acc.$\mid$incompl. \\
\midrule
PubLayNet multi-hop & baseline & 0.565 & 0.000 \\
                    & +KG      & 0.546 & 0.000 \\
\midrule
SPIQA cross-paper   & baseline & 0.847 & 0.584 \\
                    & +KG      & 0.865 & 0.640 \\
                    & +KGret   & 0.887 & 0.776 \\
\midrule
HotpotQA bridge     & baseline & 0.977 & 0.346 \\
                    & +KG      & 0.977 & 0.372 \\
                    & +KGret   & 0.913 & 0.435 \\
\midrule
HotpotQA comparison & baseline & 0.953 & 0.714 \\
                    & +KG      & 0.948 & 0.679 \\
                    & +KGret   & 0.939 & 0.500 \\
\bottomrule
\end{tabular}
\end{table}

Accuracy on incomplete-evidence questions differs sharply by corpus. It is 0 on
PubLayNet multi-hop questions, but reaches 0.584 on SPIQA cross-paper questions
and 0.346--0.714 on HotpotQA. These answers cannot be attributed to complete
gold evidence in the retrieved context and are consistent with prior knowledge
or other non-gold contextual cues. Raw accuracy therefore overstates the amount
of performance attributable to retrieval on widely disseminated corpora.

The +KGret result is not explained by this floor. On SPIQA, its advantage
persists within the complete-evidence stratum (0.887 versus 0.847 for the
baseline). On HotpotQA, changes in conditional means should be interpreted with
care because +KGret moves harder questions from the incomplete to the complete
stratum. Reporting conditioned accuracy alongside raw accuracy makes this
composition effect visible.

\begin{figure}[t]
\centering
\includegraphics[width=\columnwidth]{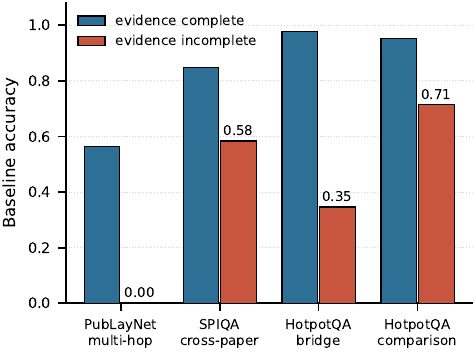}
\caption{Baseline accuracy conditioned on evidence completeness. On
PubLayNet, questions without complete gold evidence are never answered
correctly; on the corpora built from prominent documents, 0.35--0.71 of such
questions are answered from parametric memory.}
\label{fig:leakage}
\end{figure}

\subsection{Efficiency}
\label{sec:efficiency}

\begin{table*}[t]
\caption{Input tokens and cost on pixel-only figure questions. Costs are in USD
per 100 questions.}
\label{tab:efficiency}
\centering
\footnotesize
\setlength{\tabcolsep}{4pt}
\resizebox{\textwidth}{!}{%
\begin{tabular}{llcccccccc}
\hline
& & \multicolumn{4}{c}{\textbf{Input tokens per question}} & \multicolumn{4}{c}{\textbf{Cost per 100 questions}} \\
Generator & Access & base. & +KG & +mm & +both & base. & +KG & +mm & +both \\
\hline
GPT-4o-mini           & closed-weight & 485 & 511 & 9477 & 9507 & 0.0075 & 0.0079 & 0.1429 & 0.1435 \\
Gemini 3.1 Flash-Lite & closed-weight & 517 & 545 & 1601 & 1633 & 0.0055 & 0.0058 & 0.0171 & 0.0172 \\
Llama 4 Scout         & open-weight   & 484 & 508 & \textbf{861} & 889 & 0.0041 & 0.0048 & \textbf{0.0084} & 0.0097 \\
Llama 4 Maverick      & open-weight   & 484 & 508 & \textbf{861} & 889 & 0.0089 & 0.0111 & 0.0189 & 0.0195 \\
\hline
GPT-4o (diagnostic)   & closed-weight & 485 & 511 & 760 & 790 & 0.1252 & 0.1317 & 0.2200 & 0.2291 \\
\hline
\end{tabular}}
\end{table*}

\begin{figure}
\centering
\includegraphics[width=\columnwidth]{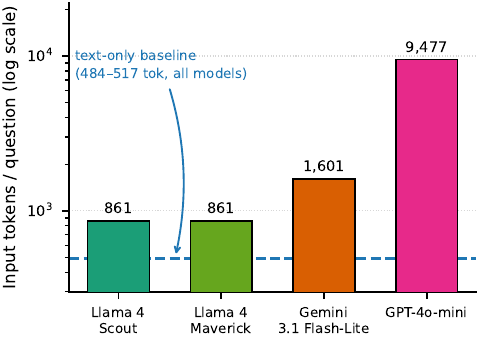}
\caption{Input tokens per question for the multimodal configuration. All models
receive the same text and image inputs.}
\label{fig:image_tokens}
\end{figure}

Multimodal cost is dominated by provider-specific image tokenisation. For the
same text and image inputs, the multimodal prompt ranges from 861 input tokens
for the Llama~4 models to 9{,}477 for GPT-4o-mini. Llama~4~Scout therefore costs
\$0.0084 per 100 pixel-only questions with accuracy 0.114, whereas GPT-4o-mini
costs \$0.1429 with accuracy 0.086. Latency does not follow the same ordering:
Gemini~3.1~Flash-Lite is fastest at 440--876\,ms, while Llama~4~Maverick reaches
1{,}876--4{,}087\,ms despite sharing Scout's input-token count.

By comparison, +KGret adds approximately 170--200 input tokens per question on
the cross-paper and bridge sets. Where an evidence deficit exists, it provides
the largest accuracy improvement per unit of added context among the tested
augmentations. These measurements show that multimodal deployment decisions
should account for image tokenisation and serving behaviour, not only listed
per-token prices.

\section{Discussion}
\label{sec:discussion}

The controlled comparisons support three conclusions about augmentation in
document RAG.

\subsection{Graph Benefit Depends on Stage and Evidence Deficit}

The strongest result is the asymmetry between generation-stage and retrieval-
stage graph use. Provenance-constrained triple injection does not produce a
consistent gain on any corpus and often reduces faithfulness. The same graph,
when used to recover entity-linked passages that text similarity missed,
increases evidence completeness and improves accuracy for every generator on
the two evidence-deficient sets. The HotpotQA bridge/comparison control further
shows that this improvement is negligible when retrieval is already nearly
complete. Graph augmentation should therefore be understood as a retrieval
intervention whose value is conditional on missing evidence, rather than as a
generic prompt enrichment strategy. This interpretation reconciles the present
results with retrieval-integrated systems such as HippoRAG~\cite{hipporag} and
with reports that plain RAG and GraphRAG are complementary across question
types~\cite{ragvsgraphrag}.

The mechanism is also comparatively simple. Most of the gain is achieved by
one-hop entity-to-provenance association rather than deep graph traversal. The
remaining ceiling arises from limited bridging capacity and unmatched entity
surface forms. This suggests that entity canonicalisation and candidate ranking
are higher-priority improvements than increasing graph depth for the current
architecture.

\subsection{Multimodal Gains Depend on Benchmark Integrity}

The matched figure protocols show that apparent multimodal value depends on
whether the answer is genuinely unavailable through text. Caption-derived
questions admit a text-only baseline because the caption itself often contains
the reference answer. The SPIQA analysis identifies an additional leakage path
through retrievable paper prose. A third source is visible when strict-pixel
questions are answered correctly despite incomplete gold evidence, indicating
that model prior knowledge or non-gold context can create a non-zero text-only
floor. Consequently, a visual question should not be labelled pixel-only solely
from its authoring instruction. The claim should be verified against both the
caption and the exact text corpus supplied to retrieval, and results should be
reported with evidence completeness.

Multimodal access remains necessary on PubLayNet's verified pixel-only set, but
it is not sufficient for reliable QA. Recall@1 limits the current visual path to
0.229 before generation is considered, and all production generators remain
well below that ceiling. The small improvement from the higher-cost diagnostic
model further indicates that retrieval granularity and figure reading are the
primary constraints.

\subsection{Evidence-Conditioned Accuracy Improves Attribution}

Correct answers without complete gold evidence are absent on PubLayNet but
substantial on SPIQA and HotpotQA. Regardless of whether this behaviour reflects
memorised facts, partial contextual cues, or both, it cannot be credited to
complete retrieved evidence. Raw end-to-end accuracy therefore conflates
retrieval quality with model prior knowledge on familiar public corpora.
Conditioning accuracy on evidence completeness provides a simple diagnostic for
this confound and makes configuration comparisons more interpretable. In this
study, the retrieval-stage graph gain remains visible after conditioning,
whereas corpus-level differences in non-retrieval-attributable accuracy become
explicit rather than hidden inside the headline score.

\subsection{Practical Implications}

The results favour selective augmentation. Graph bridging is useful when a query
requires entity-linked evidence that dense retrieval fails to recover; otherwise
it adds context with little benefit. Visual input should likewise be gated to
questions with a genuine visual evidence requirement. Deployment cost must be
measured on the actual serving stack because the same image produces an
elevenfold token difference across providers, while latency follows a separate
ordering. These observations argue against treating either graph or multimodal
augmentation as an unconditional architectural upgrade.

\subsection{Limitations and Future Work}
\label{sec:limitations}

Question sets contain 26--50 items per condition, so small differences correspond
to few examples. Claims are therefore based primarily on consistent effects
across generators and matched controls rather than isolated cells. Hosted model
endpoints also introduce run-to-run variation, and binary LLM judges remain
subject to known evaluation biases~\cite{zheng2023judge}. Human validation of a
representative subset would strengthen calibration.

The caption-answerable protocol is evaluated only on PubLayNet, while the
strict-pixel, stage, and evidence-conditioning analyses span multiple corpora.
SPIQA and HotpotQA consist of widely disseminated public documents, which makes
non-retrieval answer generation plausible but does not establish the exact
contents of any model's pretraining data. The open-weight models are evaluated
through hosted inference, so the study does not assess the operational benefits
of local deployment.

Future work should evaluate entity canonicalisation and coreference for graph
bridging, hybrid visual retrievers that exploit OCR or region-level structure
for numerical figures and tables, and larger human-validated question sets over
connected document collections. The integrity checks and evidence-conditioned
reporting used here are independent of the specific retrievers and can be
applied to broader document-RAG benchmarks.

\section{Conclusion}

This study shows that the value of graph and multimodal augmentation in document
RAG is conditional on how evidence enters the pipeline and on how evaluation
questions are constructed. The same knowledge graph is ineffective when used
only to inject reformulated facts, yet consistently beneficial when it recovers
missing passages for evidence-deficient queries. Multimodal access is necessary
for verified pixel-only questions, but measured gains are inflated when answers
remain recoverable from captions or corpus text. Finally, substantial accuracy
without complete retrieved evidence on widely disseminated corpora demonstrates
that raw accuracy alone cannot identify retrieval-attributable progress.
Stage-controlled ablations, integrity verification, and evidence-conditioned
reporting therefore provide a more defensible basis for evaluating augmented
RAG systems.


\section*{CRediT authorship contribution statement}
\textbf{Sokipriala Jonah:}
Conceptualization, Data curation, Formal analysis, Investigation,
Methodology, Software, Writing, review and editing.

\textbf{Opeoluwa Akinseloyin:}
Validation, Writing, review and editing.

\textbf{Michael Ajao-Olarinoye:}
Software, Writing, review and editing.

\textbf{Abiola Babatunde:}
Software, Writing, review and editing.

\section*{Declaration of Competing Interest}
The authors declare that they have no known competing financial interests
or personal relationships that could have appeared to influence the work
reported in this paper.

\section*{Declaration of Generative AI and AI-Assisted Technologies in the
Writing Process}
During the preparation of this work, the authors used Claude to support
writing refinement and selected coding tasks. All outputs were reviewed and
edited by the authors, who take full responsibility for the content of the
article.

\section*{Data Availability}
The question sets, per-question run logs, generated captions, knowledge-graph
triple caches, ingestion scripts that reconstruct each corpus from its public
distribution, and the scripts that regenerate every table and figure in this
article are available at
\url{ https://github.com/sokistar24/multimodal-graph-rag}.

\section*{Acknowledgements}
The authors acknowledge the Centre for Computational Science and
Mathematical Modelling at Coventry University for providing the research
environment that supported the development of this work.

\bibliographystyle{elsarticle-num}
\bibliography{references}

@inproceedings{lewis2020rag,
  author    = {Patrick Lewis and Ethan Perez and Aleksandra Piktus and Fabio Petroni and Vladimir Karpukhin and Naman Goyal and Heinrich K{\"u}ttler and Mike Lewis and Wen-tau Yih and Tim Rockt{\"a}schel and Sebastian Riedel and Douwe Kiela},
  title     = {Retrieval-Augmented Generation for Knowledge-Intensive {NLP} Tasks},
  booktitle = {Advances in Neural Information Processing Systems (NeurIPS)},
  volume    = {33},
  pages     = {9459--9474},
  year      = {2020}
}

@inproceedings{karpukhin2020dpr,
  author    = {Vladimir Karpukhin and Barlas O{\u{g}}uz and Sewon Min and Patrick Lewis and Ledell Wu and Sergey Edunov and Danqi Chen and Wen-tau Yih},
  title     = {Dense Passage Retrieval for Open-Domain Question Answering},
  booktitle = {Proceedings of the 2020 Conference on Empirical Methods in Natural Language Processing (EMNLP)},
  pages     = {6769--6781},
  year      = {2020}
}

@article{gao2023survey,
  author  = {Yunfan Gao and Yun Xiong and Xinyu Gao and Kangxiang Jia and Jinliu Pan and Yuxi Bi and Yi Dai and Jiawei Sun and Meng Wang and Haofen Wang},
  title   = {Retrieval-Augmented Generation for Large Language Models: A Survey},
  journal = {arXiv preprint arXiv:2312.10997},
  year    = {2023}
}

@inproceedings{yang2024crag,
  author    = {Xiao Yang and Kai Sun and Hao Xin and Yushi Sun and Nikita Bhalla and Xiangsen Chen and Sajal Choudhary and Rongze Daniel Gui and Ziran Will Jiang and Ziyu Jiang and Lingkun Kong and Brian Moran and Jiaqi Wang and Yifan Ethan Xu and An Yan and Chenyu Yang and Eting Yuan and Hanwen Zha and Nan Tang and Lei Chen and Nicolas Scheffer and Yue Liu and Nirav Shah and Rakesh Wanga and Anuj Kumar and Wen-tau Yih and Xin Luna Dong},
  title     = {{CRAG} -- Comprehensive {RAG} Benchmark},
  booktitle = {Advances in Neural Information Processing Systems (NeurIPS), Datasets and Benchmarks Track},
  year      = {2024}
}

@inproceedings{ragas,
  author    = {Shahul Es and Jithin James and Luis Espinosa-Anke and Steven Schockaert},
  title     = {{RAGAS}: Automated Evaluation of Retrieval Augmented Generation},
  booktitle = {Proceedings of the 18th Conference of the European Chapter of the Association for Computational Linguistics: System Demonstrations},
  pages     = {150--158},
  year      = {2024}
}

@inproceedings{xia2025mmedrag,
  author    = {Peng Xia and Kangyu Zhu and Haoran Li and Tianze Wang and Weijia Shi and Sheng Wang and Linjun Zhang and James Zou and Huaxiu Yao},
  title     = {{MMed-RAG}: Versatile Multimodal {RAG} System for Medical Vision Language Models},
  booktitle = {International Conference on Learning Representations (ICLR)},
  year      = {2025}
}

@article{wang2025pixelrag,
  author  = {Yichuan Wang and Zhifei Li and Zirui Wang and Paul Teiletche and Lesheng Jin and Matei Zaharia and Joseph E. Gonzalez and Sewon Min},
  title   = {{PixelRAG}: Web Screenshots Beat Text for Retrieval-Augmented Generation},
  journal = {arXiv preprint},
  year    = {2025},
  note    = {TODO: add arXiv identifier / final venue once confirmed}
}

@article{faysse2024colpali,
  author  = {Manuel Faysse and Hugues Sibille and Tony Wu and Bilel Omrani and Gautier Viaud and C{\'e}line Hudelot and Pierre Colombo},
  title   = {{ColPali}: Efficient Document Retrieval with Vision Language Models},
  journal = {arXiv preprint arXiv:2407.01449},
  year    = {2024}
}

@article{edge2024graphrag,
  author  = {Darren Edge and Ha Trinh and Newman Cheng and Joshua Bradley and Alex Chao and Apurva Mody and Steven Truitt and Jonathan Larson},
  title   = {From Local to Global: A Graph {RAG} Approach to Query-Focused Summarization},
  journal = {arXiv preprint arXiv:2404.16130},
  year    = {2024}
}

@inproceedings{hagberg2008networkx,
  author    = {Aric A. Hagberg and Daniel A. Schult and Pieter J. Swart},
  title     = {Exploring Network Structure, Dynamics, and Function using {NetworkX}},
  booktitle = {Proceedings of the 7th Python in Science Conference (SciPy)},
  pages     = {11--15},
  year      = {2008}
}

@inproceedings{radford2021clip,
  author    = {Alec Radford and Jong Wook Kim and Chris Hallacy and Aditya Ramesh and Gabriel Goh and Sandhini Agarwal and Girish Sastry and Amanda Askell and Pamela Mishkin and Jack Clark and Gretchen Krueger and Ilya Sutskever},
  title     = {Learning Transferable Visual Models from Natural Language Supervision},
  booktitle = {Proceedings of the 38th International Conference on Machine Learning (ICML)},
  pages     = {8748--8763},
  year      = {2021}
}

@article{johnson2019faiss,
  author  = {Jeff Johnson and Matthijs Douze and Herv{\'e} J{\'e}gou},
  title   = {Billion-Scale Similarity Search with {GPUs}},
  journal = {IEEE Transactions on Big Data},
  volume  = {7},
  number  = {3},
  pages   = {535--547},
  year    = {2019}
}

@inproceedings{zhong2019publaynet,
  author    = {Xu Zhong and Jianbin Tang and Antonio Jimeno Yepes},
  title     = {{PubLayNet}: Largest Dataset Ever for Document Layout Analysis},
  booktitle = {Proceedings of the International Conference on Document Analysis and Recognition (ICDAR)},
  pages     = {1015--1022},
  year      = {2019}
}

@misc{lhoestq_small_publaynet_wds,
  author       = {{lhoestq}},
  title        = {{small-publaynet-wds}},
  howpublished = {\url{https://huggingface.co/datasets/lhoestq/small-publaynet-wds}},
  note         = {Hugging Face dataset, accessed 18 July 2026}
}

@misc{openai2024embeddings,
  author       = {{OpenAI}},
  title        = {New Embedding Models and API Updates},
  year         = {2024},
  month        = jan,
  howpublished = {\url{https://openai.com/index/new-embedding-models-and-api-updates/}},
  note         = {Accessed 18 July 2026}
}

@article{cho2024m3docrag,
  author  = {Cho, Jaemin and Mahata, Debanjan and Irsoy, Ozan and
             He, Yujie and Bansal, Mohit},
  title   = {{M3DocRAG}: Multi-modal Retrieval Is What You Need for
             Multi-page Multi-document Understanding},
  journal = {arXiv preprint arXiv:2411.04952},
  year    = {2024},
  doi     = {10.48550/arXiv.2411.04952}
}

@article{dong2025mmdocrag,
  author  = {Dong, Kuicai and Chang, Yujing and Huang, Shijie and
             Wang, Yasheng and Tang, Ruiming and Liu, Yong},
  title   = {Benchmarking Retrieval-Augmented Multimodal Generation for
             Document Question Answering},
  journal = {arXiv preprint arXiv:2505.16470},
  year    = {2025},
  doi     = {10.48550/arXiv.2505.16470},
  note    = {Accepted to the NeurIPS 2025 Datasets and Benchmarks Track}
}

@article{wang2026mkgragbench,
  author  = {Wang, Xiaochen and Hoang, Bao and Liu, Han and
             Wang, Ting and Ma, Fenglong},
  title   = {{MKG-RAG-Bench}: Benchmarking Retrieval in Multimodal
             Knowledge Graph-Augmented Generation},
  journal = {arXiv preprint arXiv:2606.26458},
  year    = {2026},
  doi     = {10.48550/arXiv.2606.26458},
  note    = {Accepted at KDD 2026}
}

@article{wang2026multimodalgraphrag,
  author  = {Wang, Yi-Cheng and Chen, Chu-Song},
  title   = {Multimodal Graph {RAG} for Long-Range Visually Rich
             Document Understanding},
  journal = {arXiv preprint arXiv:2606.28780},
  year    = {2026},
  doi     = {10.48550/arXiv.2606.28780}
}

@inproceedings{hipporag,
  author    = {Guti{\'e}rrez, Bernal Jim{\'e}nez and Shu, Yiheng and Gu, Yu and Yasunaga, Michihiro and Su, Yu},
  title     = {{HippoRAG}: Neurobiologically Inspired Long-Term Memory for Large Language Models},
  booktitle = {Advances in Neural Information Processing Systems (NeurIPS)},
  year      = {2024},
  note      = {arXiv:2405.14831}
}

@article{ragvsgraphrag,
  author  = {Han, Haoyu and Shomer, Harry and Wang, Yu and Lei, Yongjia and Guo, Kai and Hua, Zhigang and Long, Bo and Liu, Hui and Tang, Jiliang},
  title   = {{RAG} vs. {GraphRAG}: A Systematic Evaluation and Key Insights},
  journal = {arXiv preprint arXiv:2502.11371},
  year    = {2025}
}

@article{guo2024lightrag,
  author  = {Guo, Zirui and Xia, Lianghao and Yu, Yanhua and Ao, Tu and Huang, Chao},
  title   = {{LightRAG}: Simple and Fast Retrieval-Augmented Generation},
  journal = {arXiv preprint arXiv:2410.05779},
  year    = {2024}
}

@inproceedings{sarthi2024raptor,
  author    = {Sarthi, Parth and Abdullah, Salman and Tuli, Aditi and Khanna, Shubh and Goldie, Anna and Manning, Christopher D.},
  title     = {{RAPTOR}: Recursive Abstractive Processing for Tree-Organized Retrieval},
  booktitle = {International Conference on Learning Representations (ICLR)},
  year      = {2024},
  note      = {arXiv:2401.18059}
}

@inproceedings{sun2024thinkongraph,
  author    = {Sun, Jiashuo and Xu, Chengjin and Tang, Lumingyuan and Wang, Saizhuo and Lin, Chen and Gong, Yeyun and Ni, Lionel M. and Shum, Heung-Yeung and Guo, Jian},
  title     = {Think-on-Graph: Deep and Responsible Reasoning of Large Language Model on Knowledge Graph},
  booktitle = {International Conference on Learning Representations (ICLR)},
  year      = {2024},
  note      = {arXiv:2307.07697}
}

@inproceedings{baek2023kaping,
  author    = {Baek, Jinheon and Aji, Alham Fikri and Saffari, Amir},
  title     = {Knowledge-Augmented Language Model Prompting for Zero-Shot Knowledge Graph Question Answering},
  booktitle = {Proceedings of the 1st Workshop on Natural Language Reasoning and Structured Explanations (NLRSE)},
  year      = {2023},
  note      = {arXiv:2306.04136}
}

@inproceedings{jiang2023structgpt,
  author    = {Jiang, Jinhao and Zhou, Kun and Dong, Zican and Ye, Keming and Zhao, Wayne Xin and Wen, Ji-Rong},
  title     = {{StructGPT}: A General Framework for Large Language Model to Reason over Structured Data},
  booktitle = {Proceedings of the 2023 Conference on Empirical Methods in Natural Language Processing (EMNLP)},
  year      = {2023},
  note      = {arXiv:2305.09645}
}

@article{peng2024graphragsurvey,
  author  = {Peng, Boci and Zhu, Yun and Liu, Yongchao and Bo, Xiaohe and Shi, Haizhou and Hong, Chuntao and Zhang, Yan and Tang, Siliang},
  title   = {Graph Retrieval-Augmented Generation: A Survey},
  journal = {arXiv preprint arXiv:2408.08921},
  year    = {2024}
}

@inproceedings{yang2018hotpotqa,
  author    = {Yang, Zhilin and Qi, Peng and Zhang, Saizheng and Bengio, Yoshua and Cohen, William W. and Salakhutdinov, Ruslan and Manning, Christopher D.},
  title     = {{HotpotQA}: A Dataset for Diverse, Explainable Multi-hop Question Answering},
  booktitle = {Proceedings of the 2018 Conference on Empirical Methods in Natural Language Processing (EMNLP)},
  pages     = {2369--2380},
  year      = {2018}
}

@article{trivedi2022musique,
  author  = {Trivedi, Harsh and Balasubramanian, Niranjan and Khot, Tushar and Sabharwal, Ashish},
  title   = {{MuSiQue}: Multihop Questions via Single-hop Question Composition},
  journal = {Transactions of the Association for Computational Linguistics},
  volume  = {10},
  pages   = {539--554},
  year    = {2022}
}

@inproceedings{ho20202wiki,
  author    = {Ho, Xanh and Nguyen, Anh-Khoa Duong and Sugawara, Saku and Aizawa, Akiko},
  title     = {Constructing a Multi-hop {QA} Dataset for Comprehensive Evaluation of Reasoning Steps},
  booktitle = {Proceedings of the 28th International Conference on Computational Linguistics (COLING)},
  pages     = {6609--6625},
  year      = {2020}
}

@inproceedings{trivedi2023ircot,
  author    = {Trivedi, Harsh and Balasubramanian, Niranjan and Khot, Tushar and Sabharwal, Ashish},
  title     = {Interleaving Retrieval with Chain-of-Thought Reasoning for Knowledge-Intensive Multi-Step Questions},
  booktitle = {Proceedings of the 61st Annual Meeting of the Association for Computational Linguistics (ACL)},
  year      = {2023},
  note      = {arXiv:2212.10509}
}

@inproceedings{asai2024selfrag,
  author    = {Asai, Akari and Wu, Zeqiu and Wang, Yizhong and Sil, Avirup and Hajishirzi, Hannaneh},
  title     = {Self-{RAG}: Learning to Retrieve, Generate, and Critique through Self-Reflection},
  booktitle = {International Conference on Learning Representations (ICLR)},
  year      = {2024},
  note      = {arXiv:2310.11511}
}

@inproceedings{roberts2020knowledge,
  author    = {Roberts, Adam and Raffel, Colin and Shazeer, Noam},
  title     = {How Much Knowledge Can You Pack Into the Parameters of a Language Model?},
  booktitle = {Proceedings of the 2020 Conference on Empirical Methods in Natural Language Processing (EMNLP)},
  pages     = {5418--5426},
  year      = {2020}
}

@inproceedings{mallen2023popqa,
  author    = {Mallen, Alex and Asai, Akari and Zhong, Victor and Das, Rajarshi and Khashabi, Daniel and Hajishirzi, Hannaneh},
  title     = {When Not to Trust Language Models: Investigating Effectiveness of Parametric and Non-Parametric Memories},
  booktitle = {Proceedings of the 61st Annual Meeting of the Association for Computational Linguistics (ACL)},
  year      = {2023},
  note      = {arXiv:2212.10511}
}

@inproceedings{kandpal2023longtail,
  author    = {Kandpal, Nikhil and Deng, Haikang and Roberts, Adam and Wallace, Eric and Raffel, Colin},
  title     = {Large Language Models Struggle to Learn Long-Tail Knowledge},
  booktitle = {Proceedings of the 40th International Conference on Machine Learning (ICML)},
  year      = {2023},
  note      = {arXiv:2211.08411}
}

@inproceedings{spiqa,
  author    = {Pramanick, Shraman and Chellappa, Rama and Venugopalan, Subhashini},
  title     = {{SPIQA}: A Dataset for Multimodal Question Answering on Scientific Papers},
  booktitle = {Advances in Neural Information Processing Systems (NeurIPS), Datasets and Benchmarks Track},
  year      = {2024},
  note      = {arXiv:2407.09413}
}

@inproceedings{mathew2021docvqa,
  author    = {Mathew, Minesh and Karatzas, Dimosthenis and Jawahar, C. V.},
  title     = {{DocVQA}: A Dataset for {VQA} on Document Images},
  booktitle = {Proceedings of the IEEE/CVF Winter Conference on Applications of Computer Vision (WACV)},
  pages     = {2200--2209},
  year      = {2021}
}

@inproceedings{masry2022chartqa,
  author    = {Masry, Ahmed and Long, Do Xuan and Tan, Jia Qing and Joty, Shafiq and Hoque, Enamul},
  title     = {{ChartQA}: A Benchmark for Question Answering about Charts with Visual and Logical Reasoning},
  booktitle = {Findings of the Association for Computational Linguistics: ACL 2022},
  pages     = {2263--2279},
  year      = {2022}
}

@inproceedings{zheng2023judge,
  author    = {Zheng, Lianmin and Chiang, Wei-Lin and Sheng, Ying and Zhuang, Siyuan and Wu, Zhanghao and Zhuang, Yonghao and Lin, Zi and Li, Zhuohan and Li, Dacheng and Xing, Eric P. and Zhang, Hao and Gonzalez, Joseph E. and Stoica, Ion},
  title     = {Judging {LLM}-as-a-Judge with {MT}-Bench and Chatbot Arena},
  booktitle = {Advances in Neural Information Processing Systems (NeurIPS), Datasets and Benchmarks Track},
  year      = {2023},
  note      = {arXiv:2306.05685}
}

\appendix

\section{Example Questions}
\label{app:questions}

Table~\ref{tab:example-questions} lists representative questions from each
category, reproduced verbatim from the released question sets together with
their reference answers and gold source pages. The examples illustrate the
construction constraints described in Section~\ref{sec:qsets}: each
question identifies the relevant variable, cohort, or measurement, and none
contains a modality cue (\emph{figure}, \emph{table}, \emph{chart}) that would
reveal which retrieval branch should be used.

\begin{table*}[t]
\centering
\caption{Representative questions from each category, drawn verbatim from the
released evaluation sets. Sources are PubLayNet page identifiers; figure
sources additionally name the cropped region.}
\label{tab:example-questions}
\renewcommand{\arraystretch}{1.25}
\begin{tabular}{@{}p{0.10\textwidth} p{0.44\textwidth} p{0.22\textwidth} p{0.12\textwidth}@{}}
\toprule
\textbf{Category} & \textbf{Question} & \textbf{Reference answer} & \textbf{Source} \\
\midrule

\textbf{Text}
& What vaccine is currently the only one used in China for leptospirosis?
& multivalent inactivated vaccine
& PMC5985562\_00009 \\[2pt]

& What is the name of the instrument developed at Laval University to measure
  doctor--patient communication competency in medical students?
& Doctor-Patient Communication Competency for Medical Students (DPCC-MS)
& PMC5658909\_00001 \\
\midrule

\textbf{Multi-hop}
& What is the odds ratio for headache in individuals with chronic pain, and
  what is it for back pain?
& 1.83 (1.36--2.46) for chronic pain; 2.72 (1.73--4.29) for back pain
& PMC4913715\_00002 \\[2pt]

& What is the 30-month overall survival rate for all CLL patients treated with
  ibrutinib, and what is it for the del(17p) subset?
& 79\% for all patients; 65\% for the del(17p) subset
& PMC5679049\_00002 \\
\midrule

\textbf{Figure,}
& What are the statistical methods used to estimate incremental cost,
  incremental QALYs, and incremental net benefit (INB)?
& ITT, two-stage least squares, three-stage least squares, uBN, uBGN, and BFL
& PMC5763423\_00016 \newline \_table0 \\[2pt]
\textbf{caption-}
& How many genes are included in the Retinitis Pigmentosa panel?
& 132 genes
& PMC5592712\_00002 \newline \_table0 \\
\textbf{answerable} & & & \\
\midrule

\textbf{Figure,}
& What is the residual deviance value for the model with 2 degrees of freedom
  in 2013?
& 44.45
& PMC5710651\_00006 \newline \_table0 \\[2pt]
\textbf{pixel-only}
& Which cell type shows the highest relative expression of LILRB3 across all
  tissues examined?
& Macrophage
& PMC4060870\_00007 \newline \_figure0 \\
\bottomrule
\end{tabular}
\end{table*}

\section{Caption-Answerable and Pixel-Only Construction}
\label{app:matched}

The caption-answerable and pixel-only figure questions are constructed from the
same crop pool (598 cropped figures and tables), the same visual retriever, and
the same production generators; they differ only in the information shown to the
question author. In the caption-answerable protocol the author sees only the
GPT-4o caption, so questions are limited to content expressible in the caption.
In the pixel-only protocol the author sees the crop itself and is instructed to
ask for information requiring direct visual inspection, such as a table entry,
axis value, or reported percentage.

The effect of this distinction is made concrete by a crop that appears in both
sets. For the alcohol-advertising content table
(\texttt{PMC5467671\_00004\_table0}), the two protocols yield questions of very
different character:

\begin{quote}
\small
\textbf{Caption-answerable.}\; \emph{What are the categories compared for the
duration and display percentages of alcohol cues and responsible drinking
statements?}\\
\textbf{Reference answer:} Drinkaware, Heineken responsibility, and Heineken.\\[4pt]
\textbf{Pixel-only.}\; \emph{What is the size responsible for drinking statements
as a percentage of total display size in the Drinktomane group?}\\
\textbf{Reference answer:} 12.98.
\end{quote}

\noindent
The caption-answerable question asks \emph{which} categories are compared, which
the one-sentence caption already states; it can therefore be answered from
retrieved caption text with no access to the image. The pixel-only question asks
for a specific cell value (12.98) that appears only in the rendered table and is
absent from the caption. This is the mechanism behind the results in
Table~\ref{tab:ceiling}: text-only systems score above zero on
caption-answerable questions but exactly zero on pixel-only questions, because
the former can be recovered as text while the latter must be read from pixels.

\section{Explainability Trace}
\label{app:trace}

For every answer the system records the evidence used, so that a reviewer can
inspect why a given response was produced. Figure~\ref{fig:trace} shows a
complete trace for a single text question under the \textbf{+both} configuration,
which is the only configuration that populates all three evidence branches
(text passages, knowledge-graph facts, and a caption). The trace reproduces the
retrieved chunk with its source page and cosine score, the graph fact injected
into the prompt, and the visual evidence considered by the caption-mediated
path.

The graph fact in this trace
(\texttt{DPCC-MS}\,$\rightarrow$\,\texttt{measures}\,$\rightarrow$\,\texttt{doctor--patient
communication}) is topically correct but restates information already present in
the retrieved passage. This is a concrete instance of the pattern reported in
Section~\ref{sec:ablation}: under provenance filtering, injected graph facts tend to
reformulate the retrieved text rather than add new evidence, which is consistent
with the absence of a reliable accuracy gain from graph augmentation on this
corpus.

\begin{figure}
\centering
\footnotesize
\fbox{\parbox{0.94\columnwidth}{%
\textbf{Question.}\; What is the name of the instrument developed at Laval
University to measure doctor--patient communication competency in medical
students?\\[3pt]
\textbf{Answer.}\; Doctor-Patient Communication Competency for Medical Students
(DPCC-MS).\\[3pt]
\textbf{Configuration.}\; +both\ \ ($\alpha{=}1,\ \beta{=}1$)\\[5pt]
\textbf{Evidence trace}\\[2pt]
\underline{Text passages (top $k{=}3$):}\\
\;[1] \texttt{PMC5658909\_00001} \; (cos $=0.71$)\\
\;\hspace{1.2em}\emph{``\ldots an instrument, based on the Patient-Centered
Clinical Method, was developed to measure doctor--patient communication
competency in medical students\ldots''}\\
\;[2] \texttt{PMC5658909\_00001} \; (cos $=0.64$)\\
\;[3] \texttt{PMC5658909\_00001} \; (cos $=0.58$)\\[3pt]
\underline{Knowledge-graph facts (whole-word match, page-filtered):}\\
\;$\bullet$\; DPCC-MS \;$\xrightarrow{\text{measures}}$\; doctor--patient communication\\
\;$\bullet$\; instrument \;$\xrightarrow{\text{is based on}}$\; Patient-Centered Clinical Method\\[3pt]
\underline{Visual evidence (caption-mediated, $\beta{=}1$):}\\
\;top-ranked crop CLIP score $=0.19$ (below use threshold; not attached)\\[3pt]
\underline{Fusion note:}\; answer supported by text passage [1]; graph facts
restate the retrieved text; no visual evidence contributed.
}}
\caption{Worked explainability trace for one text question under the +both
configuration. Chunk excerpts are abbreviated; source pages, cosine scores, the
injected graph facts, and the CLIP score are taken from the system logs.}
\label{fig:trace}
\end{figure}
%


\section{Example Questions: SPIQA and HotpotQA}
\label{app:questions_ext}

Table~\ref{tab:examples_ext} lists representative questions from the
extended-corpus evaluation sets, complementing the PubLayNet examples of
Appendix~\ref{app:questions}. Cross-paper items carry two gold papers and contain their linking
entity verbatim, as required by the seeding protocol of
Section~\ref{sec:qsets}; HotpotQA items carry their gold paragraphs as
multi-source lists.

\begin{table*}[t]
\centering
\caption{Representative questions from the SPIQA and HotpotQA evaluation
sets, drawn verbatim from the released files. Cross-paper sources list both
gold papers; the linking entity is italicised in the description column.}
\label{tab:examples_ext}
\renewcommand{\arraystretch}{1.25}
\begin{tabular}{@{}p{0.13\textwidth} p{0.42\textwidth} p{0.23\textwidth} p{0.14\textwidth}@{}}
\toprule
\textbf{Category} & \textbf{Question} & \textbf{Reference answer} & \textbf{Source(s)} \\
\midrule

\textbf{SPIQA text}
& What is the name of the algorithm proposed for mining informative (feature,
  value) pairs in coreference resolution?
& EPM
& 1708.00160v2 \\[2pt]
& What loss function is proposed for end-to-end F-measure maximization in
  salient object detection?
& FLoss
& 1805.07567v2 \\
\midrule

\textbf{SPIQA multi-hop (within-paper)}
& In the MTSA mechanism, which type of self-attention captures pairwise
  dependency, and which type captures global dependency?
& Pairwise dependency is captured by dot-product based token2token
  self-attention, and global dependency is modeled by feature-wise multi-dim
  source2token self-attention.
& 1805.00912v4 \\
\midrule

\textbf{SPIQA multi-hop (cross-paper)}
& How are pretrained word2vec embeddings utilized in the training process, and
  what effect do they have on internal coherence in models for IMDB and Yahoo
  answers? \emph{(entity: pretrained word2vec embeddings)}
& Initialized and not updated, returning the model with the best internal
  coherence for IMDB and Yahoo answers; words with count 1 are replaced with
  the UNK token with probability 0.5.
& 1804.07707v2 \newline 1705.09296v2 \\[2pt]
& How is maximum likelihood estimation applied in supervised summarization
  approaches, and what role does it play in pre-training the generator in
  GANs? \emph{(entity: maximum likelihood estimation)}
& Defines the similarity measure based on the Multi-DPP measure with respect
  to ground-truth labels; pre-trains the generator on the parallel training
  set.
& 1812.00108v4 \newline 1703.04887v4 \\
\midrule

\textbf{HotpotQA bridge}
& According to the 2001 census, what was the population of the city in which
  Kirton End is located?
& 35{,}124
& Boston\_Lincolnshire \newline Kirton\_End \\[2pt]
& Andrew Jaspan was the co-founder of what not-for-profit media outlet?
& The Conversation
& Andrew\_Jaspan \newline The\_Conversation \\
\midrule

\textbf{HotpotQA comparison}
& Were Scott Derrickson and Ed Wood of the same nationality?
& yes
& Ed\_Wood \newline Scott\_Derrickson \\
\bottomrule
\end{tabular}
\end{table*}

The first bridge item illustrates why bridge questions constitute the graph's
test case. The question names \emph{Kirton End}; the answer (a census
population) appears only in the article on \emph{Boston, Lincolnshire}, which
the question never mentions. Text retrieval ranks the Boston paragraph low
because it shares no surface terms with the question, whereas the extracted
triple linking Kirton End to Boston identifies the Boston article as a bridge
page, from which the candidate-set expansion of Eq.~\ref{eq:kgret} recovers
the required paragraph. Comparison questions such as the final item name both
entities, so both gold paragraphs are directly retrievable and the expansion
has nothing to add--the asymmetry exploited as an internal control in
Section~\ref{sec:mechanism}.

\section{Figure-Question Integrity Protocol}
\label{app:integrity}

The integrity verification of Section~\ref{sec:integrity} applies two
programmatic checks to every figure question, using normalised string
containment (lower-cased, punctuation-stripped) of the reference answer:

\begin{enumerate}
  \item \textbf{Caption check.} The answer must not appear in the crop's
  pipeline caption. A hit means the question is answerable from retrieved
  caption text with no image access.
  \item \textbf{Corpus check.} The answer must not appear in the source
  document's retrievable text. A hit means the corpus's text channel restates
  the figure content, so a text-only system can recover the answer through
  ordinary passage retrieval.
\end{enumerate}

On PubLayNet, whose text channel is page-level OCR of text regions, the
caption check yields 0 of 35 hits and the corpus check is closed by
construction, since figure content is not part of the OCR output. On SPIQA,
whose text channel is TeX-derived full-paper prose, the checks partition the
35 questions into the tiers of Table~\ref{tab:integrity}: five
caption-contaminated items, four text-recoverable items, and 26 strict-pixel
items. The five excluded crops are
\texttt{1809.00263v5-Figure12-1},
\texttt{1906.10843v1-Figure5-1},
\texttt{1705.02946v3-Figure6-1},
\texttt{1803.01128v3-Table2-1}, and
\texttt{1802.07351v2-Figure10-1}.
Text-only generators answered two of the five contaminated items and up to
one of the four text-recoverable items, confirming that both flagged leak
paths are exploitable in practice.

A third leak path is not detectable by construction-time checks: on the
strict-pixel tier, text-only accuracy is 2--3 of 26 rather than zero, and the
evidence-conditioned analysis of Section~\ref{sec:leakage} attributes this
residual to parametric memory of prominent papers. The protocol therefore
verifies what can be verified--caption and corpus recoverability--and
quantifies what cannot be closed, by pairing the verified tier with an
obscure-corpus reference point on which the same protocol yields an exact
zero. Both checks run in seconds from the released question files, captions,
and corpus, and are included in the repository so that the tier assignment of
every question is reproducible.


\end{document}